\documentclass{article}
\usepackage[dvips]{graphicx,color}
\usepackage{amsmath,amssymb}
\newcommand{\bce}{\begin{center}}
\newcommand{\ece}{\end{center}}
\newcommand{\be}{\begin{equation}}
\newcommand{\ee}{\end{equation}}
\newcommand{\bea}{\begin{eqnarray}}
\newcommand{\eea}{\end{eqnarray}}
\newcommand{\bdes}{\begin{description}}
\newcommand{\edes}{\end{description}}
\newcommand{\bit}{\begin{itemize}}
\newcommand{\eit}{\end{itemize}}
\newcommand{\btt}{\begin{tt}}

\renewcommand{\thesection}{\arabic{section}.}
\renewcommand{\thesubsection}{\thesection\arabic{subsection}}
\renewcommand{\theequation}{\arabic{equation}}

\newcommand{\E}{\> = \>}
\newcommand{\EA}{&=&}
\newcommand{\non}{\nonumber \\}
\newcommand{\no}{\nonumber}

\newcommand{\Eqi}{\> \equiv \>}
\def\fb{{\bf b}} 

\def\fk{{\bf k}}
\def\fK{{\bf K}}
\def\fp{{\bf p}}
\def\fq{{\bf q}}

\def\fu{{\bf u}}
\def\fv{{\bf v}}

\def\fx{{\bf x}}

\def\fy{{\bf y}}
\def\fz{{\bf z}}

\def\vecxi{\mbox{\boldmath$\xi$}}
\def\vecrho{\mbox{\boldmath$\rho$}}

\def\Iint{\int_{-\infty}^{+\infty} dt}

\def\Def{\> := \>}
\def\deF{\> =: \>}
\def\la{\left\langle \,}
\def\ra{\, \right\rangle}
\def\lrp{\left ( \, }    % left round parenthesis
\def\rrp{\, \right ) }   % right round parenthesis
\def\lsp{\left [ \, }    % left square parenthesis
\def\rsp{\, \right ] }   % right square parenthesis
\def\lcp{\left \{ \, }   % left curly parenthesis
\def\rcp{\, \right \} }  % right curly parenthesis
\def\lvl{\, \left | \, }     % left vertical line
\def\rvl{\, \right | \, }    % right vertical line
\begin{document}
\begin{flushright}
PSI-PR-13-02\\
\end{flushright}
\setcounter{page}{1}
\thispagestyle{empty}

\vspace{2cm}

\bce
{\Large\bf Scattering Theory with Path Integrals}\\

\vspace{1cm}

{\large R.~Rosenfelder}\\

\vspace{1cm}

\noindent
Particle Theory Group, Paul Scherrer Institute,
CH-5232 Villigen PSI, Switzerland\\

\ece

\vspace{2cm}

\begin{abstract}
\noindent
Starting from well-known expressions for the $T$-matrix and its derivative in standard 
nonrelativistic potential scattering I rederive recent path-integral formulations due to Efimov and 
Barbashov {\it et al.} Some new relations follow immediately.

\end{abstract}
\newpage

%%%%%%%%%%%%%%%%%%%%%%%%%%%%%%%%%%%%%%%%%%%%%%%%%%%%%%%%%%%%%%%%%%%%%%%
%%%  ********************* INTRODUCTION ******************************
%%%%%%%%%%%%%%%%%%%%%%%%%%%%%%%%%%%%%%%%%%%%%%%%%%%%%%%%%%%%%%%%%%%%%%

\section{Introduction}

Traditionally the path-integral method in quantum physics has been applied mostly to bound-state problems.
Following the pioneering work of Ref. \cite{CFJM} there has been renewed interest 
in the last few years to use it also for scattering problems \cite{Rose1,CaRo2,Efim} in nonrelativistic
physics. This offers the chance of finding new approximation methods \cite{Carr,CaRo1} or trying to evaluate
the involved path integrals by stochastic methods although the oscillating nature 
of the real-time
path integral presents a great challenge (see e.g. Ref. \cite{Rose2}). It should be also kept in mind that 
the path-integral approach, where one integrates functionally over the degrees of
freedom weighted by the exponential of the classical action, is 
much more general than a Schr\"odinger description thus allowing an immediate  
generalization to many-body or field-theoretical problems \cite{Schul,Klein,Pfad}. In the following, however,
only single-channel nonrelativistic scattering in a local potential $ V(\fx) $ is considered
with the aim to find practical path-integral representations for the scattering amplitude free of infinite
(time-)limits. Extensions to nonlocal potentials have been considered in Ref. \cite{Rose2}.

Most of the results so far have been derived by non-standard methods specific for path integrals. For example,
in Ref. \cite{Rose1} the $S$-matrix has been considered as infinite-time limit of the evolution operator
in the interaction picture but the extraction of the energy-conserving $\delta$-function, i.e. an expression for 
the $T$-matrix was only possible by applying the Faddeev-Popov trick and 
introducing  "anti-velocity" degrees of freedom which take away unphysical
contributions in the path integral. In the meantime path-integral formulations for the $T$-matrix have been
found (or rediscovered \cite{CaRo2}) which do not need an "anti-velocity" at all but are more involved. 
They are either based 
on a technique invented by Barbashov and collaborators in quantum field theory \cite{Barb_etal} or use the asymptotic 
limit of the path-integral solution of the Schr\"odinger equation \cite{Efim}.

One may wonder whether it is not possible to start from well-developed nonrelativistic scattering theory as 
available in many excellent text books \cite{Mess,RoTha,GoWat,Tayl,Newt}. This is the aim of the present note and, 
indeed, I succeed in deriving old -- and also some new -- path-integral expressions for the scattering amplitude which 
may be of some use for further analytical or numerical studies. To be self-contained some additional material
about path integrals and derivatives of the $T$-matrix is collected in two appendices.

\section{Final (Efimov's) form for the $T$-matrix}
Here we start from the standard expression for the $T$-matrix 
\footnote{In the following a quantum-mechanical operator 
is distinguished by a hat, 3-vectors are in bold face type  and a system of units is used where $ \hbar = 1 $ .} 
for scattering from a local potential $ V(\fx) $ (see e. g. Ref. \cite{Mess}, eq. XIX.17)
\be
T_{i \to f} \E \la \psi_f^{(-)} \lvl  \hat V \rvl \phi_i \ra
\label{T final}
\ee
where $ \> \la \fx \lvl \phi_i \right . \ra = \exp ( i \, \fk_i \cdot \fx ) \> $ is a free plane wave with initial momentum $ \fk_i $ and
\be
\la  \psi_f^{(-)} \rvl  \E \la \phi_f \rvl
\lrp \hat U_I(0,+\infty) \rrp^{\dagger}  \E \la \phi_f \rvl \hat U_I(+\infty,0) 
\ee
is the full incoming scattering wave (Ref. \cite{RoTha}, ch. 8, eq. (3.4)) with final momentum $ \fk_f$. 
The latter is obtained from the free 
wave by application of the M{\o}ller operator 
\be
\hat U_I(+\infty,0) \E  \lim_{t_1 \to \infty} e^{i \hat H_0 t_1} \, \hat U(t_1,t_0) \, e^{-i \hat H_0 t_0} \> 
\Bigr |_{t_0=0} \> ,
\label{UI inf,0}
\ee
which is the time-evolution operator in the the interaction picture associated with the free/full Hamiltonians
$ \> \hat H_0 / \hat H \> $  (Ref. \cite{RoTha}, ch. 8, eq. (1.22)). The delicate infinite-time limit can be controlled
by making the potential slightly time-dependent, e. g. by multiplying it with a factor $ \> \exp(- \epsilon |t| ) \> $
and setting $ \epsilon = 0 $ at the very end of the calculation, i.e. by switching off the interaction at 
very early and very late times. If we insert Eq. \eqref{UI inf,0} into Eq. \eqref{T final} we 
obtain
\be
T_{i \to f} \E \lim_{T \to \infty} \, e^{i E T}  \la \phi_f \lvl  \hat U(T,0) \, \hat V \rvl \phi_i \ra
\E  \lim_{T \to \infty} \, e^{i E T}  \int d^3x \, d^3y \> 
\bigl < \phi_f \, | \, \fy \bigr > \la \fy \lvl  \hat U(T,0) \rvl \fx \ra \,  V(\fx) \, \bigl <  \fx \, 
| \,  \phi_i \bigr >
\ee
where $ E = \fk_i^2/(2m) = \fk_f^2/(2m) $ is the scattering energy, and we
now can employ the path-integral representation \eqref{path xi} for the full time-evolution operator 
as presented in Appendix A. This gives
\bea
T_{i \to f} \EA \lim_{T \to \infty} \, e^{i E T} \, \int d^3x \, d^3y \> e^{-i \fk_f \cdot \fy}  \, 
 \lrp \frac{m}{2 \pi i \, T} \rrp^{3/2} \exp \lsp i \frac{m}{2} \frac{(\fy - \fx)^2}{T} \rsp \non
&& \times \, {\cal N}_{\xi} \int_{\vecxi(0)=0}^{\vecxi(T)=0} {\cal D}^3\xi \> \exp \lcp i \int_0^T dt \, 
\lsp \frac{m}{2} \dot \vecxi^2
- V \lrp \fx + \frac{\fy-\fx}{T} t + \vecxi(t) \rrp \rsp \rcp \> V(\fx) \, e^{i \fk_i \cdot \fx} \> .
\eea
The normalization $ \> {\cal N}_{\xi} \> $ (Eq. \eqref{N xi} with $ t_b - t_a = T $ ) ensures that the free 
path integral gives unity. 
After substituting  $\> \fy = \fx + \fu \, T  \> $ we can perform the $\fu $-integration in the large $T$-limit by the stationary phase method 
\be
\lrp \frac{m T}{2 \pi i} \rrp^{3/2} \, \int d^3u \> \exp \lsp i \lrp - \fk \cdot \fu + 
 \frac{m}{2} \fu^2 \rrp \, T  \rsp \, F(\fu) \> \stackrel{T \to \infty}{\simeq} \> \exp \lrp - i \frac{\fk^2}{2 m} T \rrp \> F \lrp \fu_{\rm stat} \rrp 
\label{stat phase}
\ee
with the stationary point
\be
\fu_{\rm stat} \E \frac{\fk}{m}  \quad , \qquad \fk \Eqi \fk_f
\ee
if the function $ F(\fu) $ does not vary much. Therefore all explicit $T$-dependence completely cancels out 
and the final result is
\be
\boxed{
T_{i \to f} \E \int d^3x \> e^{-i \fq \cdot \fx} \, V(\fx) \, 
\> {\cal N}_{\xi} \int_{\vecxi(0)=0}^{\vecxi(\infty)=0} {\cal D}^3\xi \> \exp \lcp i \int_0^{\infty} dt \, 
\lsp \frac{m}{2} \dot \vecxi^2
- V \lrp \fx+ \frac{\fk_f}{m}t + \vecxi(t) \rrp \rsp \rcp \> ,
\label{path f}
}
\ee
where $\> \fq = \fk_f - \fk_i \> $ denotes the momentum transfer. This 
obviously reduces to the first Born approximation if the 
path integral, i.e. higher orders, are neglected. It is also in agreement 
with the result given by Efimov \cite{Efim}~\footnote{Apart from an obvious  minus 
sign missing in his eq. (64) as can be seen by comparing eqs. (60) and (63).}
for the scattering amplitude  $ \> f_{i \to f}  = - m \, T_{i \to f}/(2 \pi) \> $.

\section{Initial form for the $T$-matrix}
There is an alternative form of the $T$-matrix in scattering theory (given e.g. in Ref. \cite{Mess}, eq. XIX.15)
\be
T_{i \to f} \E \la \phi_f \lvl \hat V \rvl \psi_i^{(+)} \ra
\label{T init}
\ee
where the full outgoing scattering wave function with momentum $ \fk_i $
\be
\lvl \psi_i^{(+)} \ra  \E \hat U_I(0,-\infty) \, \lvl \phi_i \ra 
\ee
enters (Ref. \cite{RoTha}, ch. 8, eq. (3.2)). Proceeding as in the previous section we obtain
\bea
T_{i \to f} \EA \lim_{T \to \infty} \, e^{i E T} \, \int d^3x \, d^3y \> e^{-i \fk_f \cdot \fx} \,  
V(\fx) \, 
 \lrp \frac{m}{2 \pi i \, T} \rrp^{3/2} \exp \lsp i \frac{m}{2} \frac{(\fx - \fy)^2}{T} \rsp \non
&& \times \, {\cal N}_{\xi} \int_{\vecxi(-T)=0}^{\vecxi(0)=0} {\cal D}^3\xi \> \exp \lcp i \int_{-T}^0 dt \,  
\lsp \frac{m}{2} \dot \vecxi^2
- V \lrp \fy + \frac{\fx-\fy}{T} (t+T) + \vecxi(t) \rrp \rsp \rcp \>  e^{i \fk_i \cdot \fy} \> .
\eea
Again we perform the substitution $ \> \fy = \fx - \fu \,  T\> $, find the stationary point in the $\fu$-integral 
as $ \> \fu_{\rm stat} = \fk_i/m \>  $  and obtain
\be
\boxed{
T_{i \to f} \E \int d^3x \> e^{-i \fq \cdot \fx} \, V(\fx) \, 
\> {\cal N}_{\xi} \int_{\vecxi(-\infty)=0}^{\vecxi(0)=0} {\cal D}^3\xi \> \exp \lcp i \int_{-\infty}^0 dt \, 
\lsp \frac{m}{2} \dot \vecxi^2 - V \lrp \fx+ \frac{\fk_i}{m}t + \vecxi(t) \rrp \rsp \rcp \> .
\label{path i}
}
\ee
It is seen that the transformations $ \> \fk_i \to -\fk_f, \fk_f \to -\fk_i   \> $ together with the change in the 
$ \> t \to -t \> $ in the $t$-integral turn Eq. \eqref{path i} into Eq. \eqref{path f}. Thus our path-integral
representations fulfill the property of microreversibility (Ref. \cite{Mess}, eq. XIX.21) or 
time-reversal invariance
\be
T_{(-f) \to (-i)} \E T_{i \to f} \> .
\label{micro rev}
\ee

\section{Symmetric (Barbashov's) form for the $T$-matrix}
\label{sec: Barb}

The path-integral expressions \eqref{path i} and \eqref{path f} are both exact but emphasize the initial 
and final momentum, respectively,  as do the starting points \eqref{T init} and \eqref{T final}. 
A more symmetrical form would
be useful, in particular when approximations are made. Otherwise time-reversal invariance could 
be violated. Of course, the arithmetic average of both expressions would suffice but 
this is be a rather awkward
procedure. What is needed is an expression for the $T$-matrix in which both ingoing and outgoing scattering waves 
are contained.

This is provided by an expression for the derivative of the $T$-matrix with respect to a parameter 
$\lambda$ of the potential \cite{Tiko}
\be
\frac{\partial T_{i \to f}}{\partial \lambda} \E \la \psi_f^{(-)} \lvl \frac{\partial \hat 
V_{\lambda}}{\partial \lambda} \rvl \psi_i^{(+)} \ra \> ,
\label{deriv V}
\ee
a proof of which is given in Appendix B. Writing
\be
\frac{\partial T_{i \to f}}{\partial \lambda} \E \la \phi_f \lvl \hat U_I(\infty,0) \,  
\frac{\partial \hat V_{\lambda}}{\partial \lambda} \, \hat U_I(0,-\infty) \rvl \phi_i \ra 
\E \lim_{T \to \infty} \, e^{2 i E T} \, \la \phi_f \lvl \hat U(T,0) \,  
\frac{\partial \hat V_{\lambda}}{\partial \lambda} \, \hat U(0,-T) \rvl \phi_i \ra
\ee
we now insert the path integral representation \eqref{velo} for matrix elements of the time-evolution operators
and obtain
\bea
\frac{\partial T_{i \to f}}{\partial \lambda} \EA \lim_{T \to \infty} \, e^{2 i E T} \, 
\int d^3x_0 \, d^3x_1 \, d^3x_2 \> e^{-i \fk_f \cdot \fx_2} \, \frac{\partial  V_{\lambda}(\fx_1)}{\partial \lambda} 
 \, e^{i \fk_i \cdot \fx_0} \non
&& \hspace{-0.5cm} \times \prod_{k=1}^2 {\cal N}_{v_k} \int {\cal D}^3 v_k \> \delta^{(3)} \lrp \fx_k - \fx_{k-1} - 
\int_{T_{k-1}}^{T_k} \! \! dt_k \, \fv_k \rrp \> 
\exp \lcp i \int_{T_{k-1}}^{T_k} d\tau_k \lsp  \frac{m}{2} \fv_k^2 - V\lrp \fy_k \rrp \rsp \rcp \> .
\eea
Here we have used functional integration over velocities $ \fv_k \Eqi \fv_k(t_k) , \, k = 1,2 $ which 
has the advantage 
that no boundary conditions have to be observed (see Appendix A). 
The normalizations ${\cal N}_{v_k}$ 
are again such that the free path integrals are one (see Eq. \eqref{N v}).
The arguments of the potential terms read (cf. Eq. \eqref{Bahn})
\be
\fy_k(\tau_k) \E \frac{\fx_k + \fx_{k-1}}{2} + \fx_{v_k}(\tau_k) \> , \quad k = 1,2
\ee
where 
\be
\fx_{v_k}(\tau_k) \Def \frac{1}{2} \, \lsp \int_{T_{k-1}}^{\tau_k} ds \> \fv_k(s)
- \int_{\tau_k}^{T_k} ds \> \fv_k(s) \rsp
\quad \mbox{with} \quad T_0 \Def -T \> , \> T_k = T_{k-1} + T  
\ee
are the quantum fluctuations around the mean position.
Eliminating the co-ordinates $ \fx_0 , \> \fx_2 $ by means of the two $\delta$-functions one obtains
\be
\fy_k(\tau_k) \E \fx_1 + \int_0^{\tau_k} ds \> \fv_k(s) \> , \quad \tau_k \in \lsp T_{k-1}, T_k \rsp
\ee
and
\be 
-\fk_f \cdot \fx_2 + \fk_i \cdot \fx_0 \E - \fq \cdot \fx_1 - \fk_i \cdot \int_{-T}^0 dt_1 \> \fv_1 -  
\fk_f \cdot \int_0^{T} dt_2 \> \fv_2 \> .
\ee 
If we now define
\be
\fv(t) \Def \lcp \begin{array}{ll}
                            \fv_1(t) + \fk_i/m & \quad t \in [-T,0] \\
                            \fv_2(t) + \fk_f/m & \quad t \in [0,+T] \end{array} \right. 
\ee
and extend the time interval from $ -T $ to $ + T $, 
we can concentrate
the functional integration into {\it one} velocity variable $ \fv(t)\, $ as in the proof 
of the composition law of the time-evolution operator in Appendix A. 
Writing $ \fx $ for $ \fx_1 $ we then have
\be
\boxed{
\frac{\partial T_{i \to f}}{\partial \lambda} \E
\int d^3x  \> e^{-i \fq \cdot \fx} \, \frac{\partial  V_{\lambda}(\fx)}{\partial \lambda} 
\, {\cal N}_v \int {\cal D}^3 v \>   \exp \lcp i \Iint \, \frac{m}{2} \fv^2(t) 
- i \Iint \, V \lrp \fx + \vecrho_{\rm ray}(t,\fv] \rrp \rcp 
\label{deri}
}
\ee
where \footnote{The nomenclature indicates that $ \vecrho $ is a function of $ t $, but a {\it functional} of $ \fv $.}
\be
\vecrho_{\rm ray}(t,\fv] \E \lsp \frac{\fk_i}{m} \Theta(-t) + \frac{\fk_f}{m} \Theta(t) \rsp t + 
\int_0^t d\tau \, \fv(\tau) \deF \fx_{\rm ray}(t) + \fx_{\rm quant}(t,\fv]\> .
\label{ray}
\ee
Note that again all dependence on the large time $T$ has disappeared because $ \> \fk_i^2/(2m) + \fk_f^2/(2m) = 2 E $.
It is seen that the particle mainly travels along the {\it ray} formed by the initial and final momentum while
the functional integration supplies the quantum fluctuations around that path.
\vspace{0.3cm}

As a special case of Eqs. \eqref{deriv V}, \eqref{deri} we multiply the original potential by a strength parameter 
$ \lambda \in [0,1] $. Integrating over $ \lambda $ from zero to one \footnote{Of course,  this integration could be 
performed analytically on the r.h.s. but for many applications the present form is preferable.} we immediately 
obtain from Eq. \eqref{deri}
the result of Ref. \cite{CaRo2}
\be
\boxed{
T_{i \to f} =
\int d^3x  \, e^{-i \fq \cdot \fx} \, V(\fx) \,  {\cal N}_v \int {\cal D}^3 v
\,   \exp \lsp i \Iint \, \frac{m}{2} \fv^2(t) \rsp \, \int_0^1 d\lambda \, 
\exp \lsp - i \lambda \Iint \, V \lrp \fx + \vecrho_{\rm ray}(t,\fv] \rrp \rsp 
\label{Barb}
}
\ee
which was previously derived using a technique developed by Barbashov {\it et al.} \cite{Barb_etal} 
in quantum field theory. 
Here we see that this time-reversal invariant representation follows from standard scattering theory.

As outlined in Appendix A any velocity path integral can also be written as ordinary (Feynman) path integral. 
Taking into account the relation 
\be
\fx_{\rm quant}(t,\fv] \E \int_0^t d\tau \> \fv(\tau) \E \fx_v(t) - \fx_v(0)
\ee
and applying Eq. \eqref{v int to xi int} to Eq. \eqref{Barb} one obtains
\bea
T_{i \to f} \EA \int d^3x  \> e^{-i \fq \cdot \fx} \, V(\fx) \,   \, \lim_{T \to \infty} \, 
\int d^3 z \> \lrp \frac{m}{2 \pi i T} \rrp^{3/2} \, \exp \lrp i \frac{m \fz^2}{2 T} \rrp \> 
{\cal N}_{\xi} \,
\int\limits_{\vecxi(-T/2)=0}^{\vecxi(+T/2)=0} \! \! {\cal D}^3 \xi \non
&& \times \>   \exp \lsp i \int_{-T/2}^{+T/2} dt \> \frac{m}{2} \dot \vecxi^2(t) \rsp \> 
\int_0^1 d\lambda \exp \lsp - i \lambda \int_{-T/2}^{+T/2} dt \> 
V \lrp \fx + \frac{\fz}{T} t + \tilde \vecrho_{\rm ray}(t,\vecxi] \rrp \rsp
\eea
where
\be
\tilde \vecrho_{\rm ray}(t,\vecxi] \E \fx_{\rm ray}(t) + \vecxi(t) - \vecxi(0) \> .
\ee
Substituting $ \> \fz = \fu \, T \> $ we can again evaluate the $ \fu$-integral
in the limit $ T \to \infty $ by the method of stationary phase 
as given by Eq. \eqref{stat phase}, but now with $ \> \fk = 0 , \>  \fu_{\rm stat} = 0 $. 
This leads to the new expression
\be
\boxed{
T_{i \to f} = \int d^3x  \, e^{-i \fq \cdot \fx} \, V(\fx) \, {\cal N}_{\xi} \! \! 
\int\limits_{\vecxi(-\infty)=0}^{\vecxi(+\infty)=0} \! \! {\cal D}^3 \xi
\,   \exp \left [ i \int\limits_{-\infty}^{+\infty} dt \, \frac{m}{2} \, \dot \vecxi^2(t) \right ] \, \int_0^1 d\lambda \,  
\exp \lsp - i \lambda  \int\limits_{-\infty}^{+\infty} dt \, V \left( \fx + \tilde \vecrho(t,\vecxi] \right) \rsp .
\label{path symm}
}
\ee
Compared to the velocity path integral \eqref{Barb} this Feynman path integral for the $T$-matrix has one
integration less (not a big deal for an infinite-dimensional functional integral...) but the paths have 
to obey specific boundary conditions. Therefore the jury is still out on deciding which form is more convenient -- for the analytic task of checking the second  Born approximation the path integral in Eq. \eqref{path symm} seems to be more involved than the one in Eq. \eqref{Barb} which only involves Gaussian integration. For other applications
the representation \eqref{path symm} by an ordinary path integral may offer advantages.
\vspace{0.3cm} 

Finally I will give a path-integral expression for the derivative of the $T$-matrix w.r.t. 
the momentum $ k = |\fk_i| = |\fk_f| $ starting from the symmetric expression \cite{Demk,Lenz,Tiko}
\be
\frac{\partial}{\partial k} \lrp k \, T_{i \to f} \rrp \E - \la \psi_f^{(-)} \bigl | \, 2 V(\hat \fx)  + 
\hat \fx \cdot \nabla V(\hat \fx)  \, \bigr | \psi_i^{(+)} \ra \>.
\label{k deriv}
\ee
As this relation can be derived by a similar scaling of co-ordinate and momentum as 
the virial theorem in the bound-state case 
(see Appendix B) one may call it the "scattering virial theorem".
Following the derivation of Eq. \eqref{deri} its path-integral representation can be written down immediately
\vspace{0.3cm}

\fcolorbox{black}{white}{\parbox{14.5cm}
{
\bea
\frac{\partial}{\partial k} \lrp k \, T_{i \to f} \rrp \EA - 
\int d^3x  \> e^{-i \fq \cdot \fx} \, \Bigl [ \, 2 V(\fx) + \fx \cdot \nabla V(\fx) \, \Bigr ]
\> {\cal N}_v \int {\cal D}^3 v \>   \exp \lsp i \Iint \> \frac{m}{2} \, \fv^2(t) \rsp \non
&& \hspace{5cm} \times 
\exp \lsp - i \Iint \> V \lrp \fx + \vecrho_{\rm ray}(t,\fv] \rrp \rsp \> . 
\no
\eea
}}
\vspace{-2cm}
\bea
\label{k deriv Barb}
\eea
\vspace{0.5cm}

\noindent
A similar expression could be given for the derivative of the $T$-matrix w.r.t. to the mass $m$ of the
quantum-mechanical particle starting from Eq. (13) in Ref. \cite{Tiko}.

\vspace{1cm} 

\section{Summary and outlook}

I have derived old (and new) path-integral expressions for the $T$-matrix in nonrelativistic potential 
scattering starting from
well-known representations of $T$ in standard scattering theory. In this way a recent representation due to 
Efimov \cite{Efim} in which the final momentum is emphasized has been rederived and an alternative
form has been obtained in which the particle mainly propagates along the initial momentum. Of course, if the 
full quantum fluctuations are taken into account both forms are equivalent.
Explicit time-reversal invariance requires
a formalism symmetric in initial and final state -- available from an expression for the {\it derivative} of the
$T$-matric w.r.t. to parameters of the potential or the energy. This gave the same result as previously 
\cite{CaRo2} derived from the field-theoretic work of Barbashov {\it et al.} While this result is written 
in terms of velocity path integrals I also have succeeded to give a new representation by ordinary 
Feynman path integrals.

It is tempting to choose the $S$-matrix itself
as infinite time-limit of the time-evolution operator in the interaction representation
\be
S_{i \to f} \E \lim_{T \to \infty} \la \phi_f \lvl \hat U_I(T,-T) \rvl \phi_i \ra \E \lim_{T \to \infty} \, 
e^{i (E_i + E_f) T} \, \la \phi_f \lvl \hat U(T,-T) \rvl \phi_i \ra 
\ee
which also exhibits the explicit time-reversal symmetry and perform similar steps. However, then one has to
enforce energy conservation by a Faddeev-Popov-like constraint in the path integral and -- choosing the simplest
form of this constraint \cite{Rose1} -- one ends up 
with a formulation where the particle travels along the mean momentum $ \fK = (\fk_i + \fk_f)/2 $ 
which is not on-shell since
$ \> K \equiv |\fK| = k \cos \Theta/2  \> $ with $ \> \Theta \> $ being the scattering angle. 
Therefore the ``dangerous'' time-dependent phase $ \exp (i (E_i + E_f) T) $ is not cancelled and 
one has to introduce ``phantom'' degrees
of freedom, i.e. ``anti-velocities'' to achieve that. The result 
\be
T_{i \to f} \E i \frac{K}{m}
\> \int d^2 b \> e^{- i \fq \cdot \fb } \, {\cal N}_v \, {\cal N}_w \, \int {\cal D}^3 v \, {\cal D}^d w \> 
\exp \lcp i \int_{-\infty}^{+\infty} dt \>  \frac{m}{2} \lsp \fv^2(t) -w^2(t) \rsp
\rcp \> \lcp  e^{ i \, \chi(\fb,\fv,w]} - 1 \rcp
\label{PI for T}
\ee
comes in two versions which are distinguished by the reference path along which the particle 
dominantly travels and the dimensionality $d$ of the ``anti-velocity'' $w(t)$.
In the first case the reference path is a straight-line path along the mean momentum $ \> \fK \> $
and a $(d=3)$-dimensional anti-velocity is needed
whereas in the  the second case the reference path is the same {\it ray} along 
the initial momentum  and along the final momentum which appeared in the symmetric form \eqref{Barb}
with an $(d=1)$-dimensional anti-velocity parallel to the mean momentum $ \fK $.
For more details see Ref. \cite{Rose1}.
\vspace{0.3cm}

Eq. \eqref{PI for T} looks like an {\it impact-parameter} representation of the 
$T$-matrix but it is {\it not} in a strict sense. This is because of the 
dependence of the phase $ \> \chi(\fb,\fv,w] \> $ on the kinematic variables and the 
angle-dependent factor $ K = k \cos(\Theta/2)$ in front of the impact-parameter integral: In a 
genuine impact-parameter representation all dependence on the scattering angle $\Theta$ 
should only reside in the factor $\exp ( - i \fq \cdot \fb )$ \cite{impact}.
Although the (initial, final or symmetric) path-integral formulations derived in this note may be quite useful in
many applications they are also not in the form of an impact parameter representation. This is obvious since
Eqs. \eqref{path i}, \eqref{path f} or \eqref{Barb}, \eqref{path symm} come all  as a "level 1" - representation
\cite{CaRo2} of the scattering amplitude
where one power of the potential appears in front of the path integral whereas an impact-parameter representation
would belong to the "level 0" - class.
\vspace{0.3cm}

It remains an open problem to find a  -- practical not only formal -- path-integral expression for the 
impact-parameter representation of the scattering amplitude.

\newpage

\bce
{\huge\bf Appendix}
\ece

\setcounter{section}{0}
\renewcommand{\thesection}{\Alph{section}}
\renewcommand{\thesubsection}{\thesection.\arabic{subsection}}
\renewcommand{\theequation}{\thesection.\arabic{equation}}
\newcommand{\adot}{\hspace*{-0.6cm}.\hspace*{0.2cm}}

%%%%%%%%%%%%%%%%%%% appendix A: 
\section{\adot Ordinary and velocity path integrals}
%\label{app: averages}
\setcounter{equation}{0}

Consider a quantum-mechanical particle of mass $m$ in $d$ (euclidean) dimensions under the 
influence of a general (even time-dependent) potential $V(x,t)$. 
\vspace{0.2cm}

\noindent
The standard path integral expression for the matrix element of the time evolution operator \cite{Schul}
\bea  
U  \lrp x_b,t_b;x_a,t_a \rrp \equiv \left < x_b \lvl \hat U(t_b,t_a) \rvl  x_a \right > 
\E \left < x_b \lvl   {\cal T} \exp \lsp \,- i \, \int_{t_a}^{t_b} d\tau \lrp 
\frac{\hat p^2}{2 m} + V(\hat x,\tau)  \rrp \rsp \, \rvl x_a \right > 
\eea
is obtained by slicing the time interval in $N$ pieces of size $\epsilon = (t_b - t_a)/N $,
decomposing
\be
\hat U \lrp t_b,t_a \rrp \E \prod_{k=1}^N \hat U \lrp t_k,t_{k-1} \rrp \> , 
\quad t_k = t_a + k \, \epsilon
\label{compo}
\ee 
using a 
short-time approximation for the individual factors and performing the limit $ N \to \infty $ while keeping the 
time difference $t_b - t_a $ fixed. This gives
\bea
\hspace{-1cm} U(x_b, t_b; x_a, t_a) \EA \lim_{N \to \infty} \>
\left (\frac{m}{2 \pi i \epsilon } \right )^{d N/2}
\int_{-\infty}^{+\infty} d^d x_1 \> d^d x_2 \> ... \> d^d x_{N-1} \non
&& \times \exp \left \{ i \epsilon \sum_{k=1}^{N} \left [
\frac{m}{2} \left ( \frac{x_k - x_{k-1}}{\epsilon} \right )^2 - 
V(x_k,t_k) \right ] \> \right \} \> , \quad x_0 \Eqi x_a \> , \> x_N \Eqi x_b
\label{Lagrange diskret} \\
&\equiv & \int_{x(t_a)=x_a}^{x(t_b)=x_b} {\cal D}^dx(t) \> \exp \lcp i \int_{t_a}^{t_b} dt \lsp 
\frac{m}{2} \dot x^2(t) - V(x(t),t) \rsp \rcp \> .
\label{Lagrange PI}
\eea
Note that no time-ordering operator $ \cal{T} $ is needed in the path-integral formulation since it
works with ordinary numbers.

%  new 10. 1. 13
To get rid of the cumbersome prefactors in the discretized form we 
normalize all path integrals to the corresponding path integral without interaction, 
i.e. we use
\be
\int_{x(t_a)=x_a}^{x(t_b)=x_b} {\cal D}^dx(t) \> \exp \lcp i \int_{t_a}^{t_b} dt \> 
\frac{m}{2} \dot x^2(t) \rcp \E \lsp \frac{m}{2 \pi i \, (t_b - t_a)} \rsp^{d/2} \, \exp \lsp i 
\frac{m}{2} \frac{(x_b-x_a)^2}{t_b - t_a} \rsp \deF {\cal N}_x^{-1} \> .
\label{N x}
\ee
Then
\bea
U(x_b, t_b; x_a, t_a) \EA \lsp \frac{m}{2 \pi i \, (t_b - t_a)} \rsp^{d/2} \, \exp \lsp i 
\frac{m}{2} \frac{(x_b-x_a)^2}{t_b - t_a} \rsp \non
&& \times \> {\cal N}_x \> \int_{x(t_a)=x_a}^{x(t_b)=x_b} {\cal D}^dx(t) \> 
\exp \lcp i \int_{t_a}^{t_b} dt \> \lsp 
\frac{m}{2} \dot x^2(t) - V(x(t),t) \rsp \rcp \> .
\label{PI normalized}
\eea
Often it is convenient to introduce a reference path around which one has to evaluate the 
quantum fluctuations. A natural choice would be the classical path but it is much easier to take a straigt line
connecting the initial and final point
\be
x(t) \deF x_{\rm straight}(t) + \xi(t) \> , \quad
x_{\rm straight}(t) \E x_a + \frac{x_b - x_a}{t_b - t_a} \lrp t - t_a \rrp \> .
\ee

Due to the boundary conditions $ \xi(t_a) = \xi(t_b) = 0 $, the free action does not acquire 
a term linear in $ \dot \xi(t) $ and one obtains
\bea
U(x_b, t_b; x_a, t_a) \EA  \lsp \frac{m}{2 \pi i \, (t_b - t_a)} \rsp^{d/2} \, \exp \lsp i 
\frac{m (x_b-x_a)^2}{2 \, (t_b - t_a)} \rsp \non 
&& \times \> {\cal N}_{\xi}\> \int_{\xi(t_a)=0}^{\xi(t_b)=0} \! 
{\cal D}^d\xi(t) \, \exp \lcp i \int_{t_a}^{t_b} dt \, \lsp 
\frac{m}{2} \dot \xi^2(t) - V \lrp x_{\rm straight}(t) + \xi(t),t \rrp \rsp \rcp 
\label{path xi}
\eea
where
\be
{\cal N}_{\xi}^{-1} \E  \int_{\xi(t_a)=0}^{\xi(t_b)=0} 
{\cal D}^d\xi(t) \> \exp \lcp i \int_{t_a}^{t_b} dt \> 
\frac{m}{2} \dot \xi^2(t) \rcp \E \lsp \frac{m}{2 \pi i (t_b - t_a)} \rsp^{d/2} \> .
\label{N xi}
\ee
%% end new
\vspace{0.3cm}

Still the Feynman path integrals have to obey boundary conditions which sometimes make formal
manipulations difficult. This is avoided if we switch to functional 
integration over velocities \cite{Pfad} by multiplying Eq.  \eqref{Lagrange diskret}  with
\be
1 \E  \prod_{k=1}^N \, \int \, d^d v_k \> \delta^{(d)} \left ( \frac{x_k -
x_{k-1}}{\epsilon} - v_k \right ) \E\epsilon^{d N} 
\prod_{k=1}^N \, \int \, d^d v_k \> \delta \left ( x_k - x_{k-1} 
- \epsilon v_k \right ) \> .
\ee
This allows integration over the $x_k$ ($ k = 1, \ldots N-1 $) and gives
\be
x_k = x_a + \epsilon \sum_{j=1}^k \, v_j 
\ee
or, in continuous notation
\be
x(t) \E x_a + \int_{t_0}^t d\tau \> v(\tau) \> .
\label{fluc x0}
\ee 
When integrating over the the $(N-1)$ inner points of the path there remains one 
$\delta$-function so that we have
\bea
U(x_b, t_b; x_a, t_a) \EA \lim_{N \to \infty}
\left ( \frac{\epsilon m}{2 \pi i} \right )^{\frac{d N}{2}} \!
 \int d^dv_1 \ldots d^dv_N \> \delta^{(d)} \left (
x_b - x_a - \epsilon \sum_{k=1}^N v_k \right )\non
&& \times \exp \left \{\>  i \epsilon \sum_{k=1}^N \lsp  \frac{m}{2}
v_k^2 - V \left (x_k = x_a + \epsilon \sum_{j=1}^k v_j,t_k
\right ) \, \rsp \> \right \} 
\label{Geschwind PI} \\
&& \hspace{-0.8cm} \equiv  {\cal N}_v
\int {\cal D}^d v(t) \,               
\delta^{(d)} \! \left ( x_b - x_a -\int\limits_{t_a}^{t_b} dt \, 
v(t) \right ) \,
\exp \left \{ \, i \int\limits_{t_a}^{t_b} dt \, \lsp  \frac{m}{2} 
v^2(t) -  V \lrp x(t),t \rrp \rsp  \right \}  . 
\label{velo}
\eea 
From Eq. \eqref{Geschwind PI} it is seen that the normalization of the velocity path integral is such that
\be
{\cal N}_v^{-1} \E \int {\cal D}^d v(t) \> \exp \lcp \, i \int_{t_a}^{t_b} dt \>  \frac{m}{2} 
v^2(t) \rcp \E 1 \> .
\label{N v}
\ee
We can also use the boundary condition at $ t = t_b $
\be
x_b \E x_a + \int_{t_a}^{t_b} dt \> v(t) 
\label{bc at t1}
\ee
to eliminate $x_a$ so that
\be
x(t) \E x_b -  \int_{t}^{t_b} d\tau \> v(\tau) \> .
\label{fluc x1}
\ee
Adding the two expressions and dividing by 2, the more symmetrical expression
\be
x(t) \E \frac{x_a + x_b}{2} + \frac{1}{2} \left [ \, \int_{t_a}^t d\tau \> v(\tau) 
-  \int_t^{t_b} d\tau \> v(\tau) \, \right ] \deF \frac{x_a + x_b}{2} +  x_v(t) \> .
\label{Bahn}
\ee 
is obtained.
Obviously $x_v(t) $ is the quantum fluctuation around the mean position $ (x_a + x_b)/2 $ whereas
Eqs. \eqref{fluc x0} and \eqref{fluc x1} describe the fluctuations around the points $x_a$ 
and $x_b$, respectively.
From Eq. \eqref{Bahn} we see that the quantum fluctuation $x_v(t)$ can also be 
expressed as
\be
x_v(t) \E \frac{1}{2} \, \int_{t_{\rm min}}^{t_{\rm max}} d\tau \> {\rm sgn} (t - \tau ) \, v(\tau)  \> , 
\quad t_{{\rm min}/{\rm max}} \Def {\rm  min}/{\rm max} \lrp t_a,t_b \rrp \> , \quad
t \in  \lsp t_{\rm min}, t_{\max} \rsp 
\label{def xv}
\ee
where $ \> {\rm sgn}(x) = 2 \Theta(x) - 1 \> $ is the sign function.
Note that $ \> d \, x_v(t)/dt = v(t) \> $
since $ \> d \, {\rm sgn}(t-\tau)/dt = 2 \, \delta(t - \tau) \> $. 

Comparing Eq. \eqref{velo} and Eq. \eqref{Lagrange PI}  one sees that
a given velocity path integral over a general functional $ F $
can always be written as an integral over an ordinary Feynman path integral
\be
\int {\cal D}^d v(t) \> \exp \lsp i \int_{t_a}^{t_b} dt \> \frac{m}{2} \, v^2(t) \rsp 
\> F\lsp x_v(t) \rsp  \E \int d^dz \>   
\int\limits_{x(t_a)=-z/2}^{x(t_b)=+z/2} {\cal D}^d x(t)  \>
 \exp \lsp i \int_{t_a}^{t_b} dt \> \frac{m}{2} \, \dot x^2(t)  \rsp \> 
F \lsp x(t) \rsp \> .
\ee
With a straightline reference path and the appropriate normalization factors 
\eqref{N v}, \eqref{N x} and \eqref{N xi} this becomes
\bea
&&  {\cal N}_v \, \int {\cal D}^d v(t) \> \exp \lsp i \int_{t_a}^{t_b} dt \> \frac{m}{2} \, v^2(t) \rsp 
\> F\lsp x_v(t) \rsp  \E 
\int d^dz \, \lsp \frac{m}{2 \pi i \, (t_b - t_a)} \rsp^{d/2} \, 
\exp \lsp i \frac{m}{2} \frac{z^2}{t_b - t_a} \rsp \non
&& \hspace{2cm} \times \> {\cal N}_{\xi} \! \! \int\limits_{\xi(t_a)=0}^{\xi(t_b)=0}  {\cal D}^d \xi(t)  \>\exp \lsp i \int_{t_a}^{t_b} dt \, \frac{m}{2} \, \dot \xi^2(t)  \rsp \>
F \lsp -\frac{z}{2} + \frac{z}{t_b - t_a} (t-t_a) + \xi(t) \rsp \> .
\label{v int to xi int}
\eea
\vspace{0.2cm} 

While the {\it composition law} \eqref{compo} (or ``semi-group'' property \cite{semigroup}) is built 
into the Lagrangian formulation and is 
easily proved in the discrete form by grouping the integrations into subproducts it is not immediately 
evident in the velocity path integral. Let us therefore evaluate
\bea
 \la x_2 \lvl \hat U(t_2,t_1) \, \hat U(t_1,t_0) \rvl x_0 \ra \EA
\int d^d x_1  \la x_2 \lvl \hat U(t_2,t_1) \rvl x_1 \ra  \la x_1 \lvl 
 \hat U(t_1,t_0) \rvl x_0 \ra \non
&& \hspace{-2cm} \E \int d^d x_1 \, \prod_{k=1}^2 \int {\cal D}^d v_k \,               
\delta^{(d)} \! \left (  x_k - x_{k-1} -\int\limits_{t_{k-1}}^{t_k} d\tau_k \, 
v_k(\tau_k) \right ) \non
&& \hspace{-2cm} \times \> \exp \left \{ \, i \int_{t_{k-1}}^{t_k} d\tau_k \, \lsp  \frac{m}{2} 
v_k^2(\tau_k) -  V \lrp \frac{x_k+x_{k-1}}{2} + x_{v_k}(\tau_k),\tau_k \rrp \rsp 
\, \right \}  \> .
\eea
Performing  the integration over $x_1$ with the help of the second $ \delta$-function ( $ k = 2 $ in the product) gives
\be
\int {\cal D}^d v_1 \, {\cal D}^d v_2 \> \> \delta^{(d)} \lrp y_{\delta} \rrp 
\> \exp \lrp i \, S_0 \rrp \ \exp \lcp - i \int_{t_0}^{t_1} d\tau_1 \, V(y_1,\tau_1)
-   i \int\limits_{t_1}^{t_2} d\tau_2 \, V(y_2,\tau_2) \rcp \> .
\ee
Here
\bea
y_{\delta} \EA  x_2 - x_0 - \int\limits_{t_1}^{t_0} d\tau_1 \> v_1(\tau_1) -  
\int\limits_{t_2}^{t_1} d\tau_2 \> v_2(\tau_2) \Eqi x_2  - x_0 - \int_{t_0}^{t_2} d\tau \> v(\tau) \\
S_0 \EA \int_{t_0}^{t_1} d\tau_1 \>   \frac{m}{2} 
v_1^2(\tau_1) +  \int_{t_1}^{t_2} d\tau_2 \> \frac{m}{2} v_2^2(\tau_2) \Eqi  \int_{t_0}^{t_2} d\tau \>   
\frac{m}{2} v^2(\tau) 
\eea
if we define
\be
v(\tau) \Def \lcp \begin{array}{ll}
                            v_1(\tau) & \quad \tau \in [t_0,t_1] \\
                                v_2(\tau) & \quad \tau \in [t_1,t_2] \> . \end{array} \right.  
\label{v full}
\ee
The arguments of the potential terms are
\be
y_1(\tau_1) \E \frac{x_2+x_0}{2} + x_{v_1}(\tau_1) - \frac{1}{2} \, \int_{t_1}^{t_2} d\tau \> v_2(\tau) 
\Eqi \frac{x_2+x_0}{2} + x_v(\tau_1) \> , \quad \tau_1 \in [t_0,t_1]
\ee
and, using the first $ \delta $-function $ ( k = 1 )$  for determining $x_1$ ,
\be
y_2(\tau_2) \E \frac{x_2+x_0}{2}  + x_{v_2}(\tau_2) + \frac{1}{2} \, \int_{t_0}^{t_1} d\tau \> v_1(\tau)  
\Eqi \frac{x_2+x_0}{2}  + x_v(\tau_2) \> , \quad \tau_2 \in [t_1,t_2] \> .
\ee
Thus
\bea
&&\int d^d x_1 \, \la x_2 \, \left | \hat U(t_2,t_1) \right | x_1 \ra \, \la x_1 \, 
\left | \hat U(t_1,t_0) \right | x_0 \ra \E \non
&& \int {\cal D}^d v \,               
\delta^{(d)} \lrp  x_2 - x_0 -\int_{t_0}^{t_2} d\tau \, 
v(\tau) \rrp \> \exp \lcp i \int_{t_0}^{t_2} d\tau \, \lsp  \frac{m}{2} 
v^2(\tau) -  V \lrp \frac{x_2+x_0}{2} + x_v(\tau),\tau \rrp \rsp 
\rcp \non
&& \hspace{7.2cm} \E  \la x_2 \lvl \hat U(t_2,t_0) \rvl x_0 \ra 
\eea
proving the composition law. 
{\it Unitarity} of the time-evolution operator follows if the Hamiltonian is hermitean, or -- 
in path-integral language -- if the action is real:
\be
\la x_2 \lvl \hat U(t_2,t_1) \, \hat U^{\dagger}(t_0=t_2,t_1) \rvl x_0 \ra \E
\int d^d x_1 \, \la x_2 \, \left | \hat U(t_2,t_1) \right | x_1 \ra \, 
\la x_0 \, \left | \hat U(t_2,t_1) \right | x_1 \ra^{\star} \> .
\ee
The last matrix element becomes
\bea
\la x_0 \, \lvl \hat U(t_2,t_1) \rvl x_1 \ra^{\star} \EA 
\int {\cal D}^d v \> \delta^{(d)} \lrp  x_0 - x_1 -\int_{t_1}^{t_2} d\tau \, v(\tau) \rrp \non 
&& \times \> 
\exp \lcp -i \int_{t_1}^{t_2} d\tau \, \lsp  \frac{m}{2} 
v^2(\tau) -  V^{\star} \lrp \frac{x_1+x_0}{2} + x_v(\tau),\tau \rrp \rsp \rcp \> .
\eea
and, if the potential is real ($ \> V^{\star} = V \> $), we have
\bea
\la x_0 \, \lvl \hat U(t_2,t_1) \rvl x_1 \ra^{\star} \EA 
\int {\cal D}^d v \> \delta^{(d)} \lrp  x_1 - x_0 -\int_{t_2}^{t_1} d\tau \, v(\tau) \rrp \> 
\exp \lsp  i \int_{t_2}^{t_1} d\tau \, \frac{m}{2} v^2(\tau) \rsp \non
&& \hspace{-0.5cm} \times \> \exp \lcp - i \int_{t_2}^{t_1} d\tau  \> V \lrp \frac{x_1+x_0}{2} + x_v(\tau),\tau \rrp  \rcp \Eqi \la x_1 \, \lvl \hat U(t_1,t_2) \rvl  x_0 \ra \> .
\eea
Therefore
\bea
\la x_2 \lvl \hat U(t_2,t_1) \, \hat U^{\dagger}(t_2,t_1) \rvl x_0 \ra \EA
\int d^d x_1 \, \la x_2 \, \left | \hat U(t_2,t_1) \right | x_1 \ra \, 
\la x_1 \, \left | \hat U(t_1,t_2) \right | x_0 \ra \non
\EA \la x_2 \, \lvl \hat U(t_2,t_2) \rvl x_0 \ra 
\E  \delta^{(d)} \lrp  x_2 - x_0 \rrp \Eqi \la x_2 \lvl \hat 1 \rvl x_0 \ra \> .
\label{unitarity}
\eea

\vspace{0.3cm}

\section{\adot Proof of Eqs. \eqref{deriv V} and \eqref{k deriv}}
\setcounter{equation}{0}

Here we recall the proof of these relations from standard (time-independent)  scattering theory.
Eq. \eqref{deriv V} is easily obtained by direct differentiation of Eq. \eqref{T init} written 
with the explicit form of the outgoing scattering wavefunction $ \> | \psi_i^{(+)} > \> $
\be
T_{i \to f} \E \la \phi_f \lvl \hat V_{\lambda} \, \lrp 1 + \frac{1}{E - \hat H_0 - \hat V_{\lambda} 
+ i \epsilon} \rrp \rvl \phi_i \ra
\ee 
and using
\bea
&&\frac{\partial}{\partial \lambda} \,  \frac{1}{E - \hat H_0 - \hat V_{\lambda} + i \epsilon} \E
 \frac{1}{E - \hat H_0 - \hat V_{\lambda} + i \epsilon} \, \frac{\partial \hat V_{\lambda}}{\partial \lambda}   \,  
\frac{1}{E - \hat H_0 - \hat V_{\lambda} + i \epsilon} \\
&& \la \psi_f^{(-)} \rvl \E \la \phi_f \rvl \, \lrp 1 + \hat V_{\lambda} \, \frac{1}{E - \hat H_0 - 
\hat V_{\lambda} + i \epsilon} \rrp \> .
\eea
\vspace{0.1cm}

\noindent
The energy derivative of the $T$-matrix in Eq. \eqref{k deriv} is obtained 
by an energy-dependent scaling
\be
\hat \fx \E  \lambda\, \fx' \> , \quad \hat \fp \E \frac{1}{\lambda} \, \hat \fp
\> , \qquad \lambda \E \frac{k_0}{k}
\ee
where $ k_0 $ is a fixed wave number. Therefore 
the plane waves become independent of the momentum $ \> k \> $ 
and the Hamiltonian turns into
\be
\hat H \E \frac{1}{\lambda^2} \lsp \frac{(\hat \fp')^2}{2m}  
+ \lambda^2 V (\lambda \fx') \rsp
\deF  \frac{1}{\lambda^2}\, \hat H' \> .
\ee
Thus the system evolves with $ \> \hat H' \> $ and scaled times 
%$ \> t/\lambda^2 \, \> $
in time-dependent or with a scaled, fixed energy 
%$ \> k_0^2/(2m) \> $ 
in time-independent scattering theory.
Finally, from the volume element there is an additional factor $ \lambda^3 \> $ so that
\be
\frac{k}{k_0} \, T_{i \to f} \E \int d^3x' \> \phi_f^{\star}(\fx') \, 
\lambda^2 \, V (\lambda \fx') \,\psi_i^{(+)}(\fx')
\ee
and all $k$-dependence resides now in the scaled potential 
$ \>  \lambda^2 V(\lambda\fx')\> $ for which we
can apply Eq. \eqref{k deriv}. Performing the differentiation
\be
\frac{\partial}{\partial k} \Bigl [  \, \lambda^2 V (\lambda \fx' ) \, \Bigr ]  \E  
\frac{\partial \lambda}{\partial k} \, \Bigl [ \, 
2 \lambda V (\lambda \fx') + \lambda^2  \fx' \cdot \nabla V (\lambda \fx') \, \Bigr ]  
 \E - \frac{\lambda^3}{k_0} \, \Bigl [ \,  2 V (\lambda \fx')  + 
\lambda \fx' \cdot\nabla V (\lambda \fx') \, \Bigr ]
\ee
and transforming back to the unscaled co-ordinates gives Eq. \eqref{k deriv}. For a radially symmetric
potential $ \> V\lrp r=|\fx| \rrp \> $ the (scattering) virial theorem depends on
\be
2 \, V(r) + r \, V'(r) \E \frac{1}{r} \, \frac{d}{dr} \lsp r^2 \, V(r) \rsp \> .
\label{deriv Vx}
\ee
%\vspace{0.1cm}
%
%\noindent
Some simple examples stand out: First, for a (repulsive) potential $ \> V(r) = \alpha/r^2
\> $ the r.h.s. of Eq. \eqref{deriv Vx} vanishes and therefore its
$T$-matrix (scattering amplitude) is proportional to $ 1/k $ (which is correct as its phase shifts are
energy-independent). Second, for a Coulomb potential $ \> V(r) = \alpha/r \> $ one has $ 2 V + r V'(r) = V $ and
therefore the relation
$ \> \alpha \, \partial T^{\rm Coulomb}/\partial \alpha = - \partial (k \, T^{\rm Coulomb})/\partial k \> $ holds which
tells us that the Coulomb scattering amplitude is a function of the Sommerfeld parameter
$ \> \gamma = m \alpha/k \> $ divided by $k$  (which is correct as can be seen from the 
exact expression, e.g. in Ref. \cite{Mess}, eq. B.25).

%%%%%%%%%%%%%%%%%%%%%%%%%%%%%%%%%%%%%%%%%%%%%%%%%%%%%%%%%%%%%%%%%%%%%%%%%
%%%% *************** REFERENCES *****************************************
%%%%%%%%%%%%%%%%%%%%%%%%%%%%%%%%%%%%%%%%%%%%%%%%%%%%%%%%%%%%%%%%%%%%%%%%

\end{document}